\documentclass[12pt]{article}
\usepackage{amssymb,amsmath,esint,titlesec}
\titleformat*{\section}{\normalsize\bf}
\titleformat*{\subsection}{\small\bf}

\begin{document}


\begin{titlepage}

\setlength{\baselineskip}{18pt}

                               \vspace*{0mm}

                             \begin{center}

{\LARGE\bf Power-law entropies for continuous systems and generalized operations}

                            \vspace*{25mm}

             \large\sf ANTHONY  \  \   J. \ \   CREACO \\
 
                           \vspace{2mm}           

        \normalsize\sf      Science Department, \\
                BMCC - The City University of New York, \\
             199 Chambers St., New York, NY 10007, USA.\\
                      acreaco@bmcc.cuny.edu\\
                         
                                   \vspace{15mm}

              \large\sf  NIKOLAOS \  KALOGEROPOULOS $^\dagger$\\

                            \vspace{2mm}
                            
 \normalsize\sf Center for Research and Applications \\
                                  of Nonlinear Systems \ (CRANS),    \\
   University of Patras, \  Patras 26500, \ Greece.                        \\
                    nikos.physikos@gmail.com \\
                         
                                    \end{center}

                            \vspace{25mm}

                     \centerline{\normalsize\bf Abstract}
                     
                           \vspace{3mm}
                     
\normalsize\rm\setlength{\baselineskip}{18pt} 

We present our view in a standing debate about the definition and meaning of power-law  entropies
for continuous systems. Our suggestion is that  such arguments should take into account the generalized operations  of 
addition and multiplication induced by the power-law entropies' composition properties. 
To be concrete, we highlight our view using the case of the \ $``q-"$, \ also known as ``Tsallis",  entropic functionals.  \\    

                           \vfill

                                                                         
                             \vfill

\noindent\rule{7cm}{0.2mm}\\  
   \noindent   $^\dagger$ {\footnotesize\rm Corresponding author}\\

\end{titlepage}
 
 
                                                                                \newpage                 

\rm\normalsize
\setlength{\baselineskip}{18pt}

\section{Introduction}

The proliferation of recently constructed entropic functionals \cite{Ts1, Ts-book} has brought  renewed interest in the examination of the foundations of Statistical Mechanics 
and on the formalism of  Thermodynamics, especially during the last two decades. The construction, in particular,  of entropies having a power-law form such as the 
$q$- (also known as ``Tsallis"-) \cite{Ts1, Ts-book} or the \ $\kappa$- entropies \cite{Kan}  has brought forth new problems that were until recently either ignored or were considered
 non-existent for the case of the far better-known and well-established Boltzmann/Gibbs/Shannon (BGS) entropy.\\

One of the problems that have emerged during the last decade, is whether one can even define  power-law entropies, such as the $q$-entropies, for continuous systems \cite{CAT, CRSA,
AMPBIV, BT}.  If this is possible, 
how would one justify such a formulation from the simpler and, in principle more fundamental due to quantum physics, discrete case. Reasonable arguments and proposals have been put 
forth on both sides of this debate, without however any of them being so convincing as to provide a consensus on this issue \cite{Abe1, Andresen, Abe2, BOT, LB, BL, PR1, OB1, OA, OB2}, at least not in our opinion.\\        

We believe, following the views partly  of  A. Einstein,  and partly of L. Boltzmann \cite{Gal, Cohen}, that the formulation of an entropic functional should 
depend on the underlying dynamics. The core problem is that  very few things are really known about the dynamical evolution of  systems of many degrees of freedom 
which may have some  possible physical interest. \\
 
The present work proposes that in order to resolve the debate for or against the applicability or the meaning of power-law entropies for continuous systems, 
one should consider addressing technical and conceptual points by taking into account the composition properties of the proposed entropies. This is a  point that has not been 
adequately appreciated so far, in our opinion, in the literature. An exception may be considered the attempt of \cite{BOT} which however was not successful, as pointed out in 
\cite{Abe2}.  We will make our views concrete by using the $q-$ entropies, since it is the simplest and most studied case 
among power-law entropic functionals.\\  

In Section 2, we present some background and our arguments and views on this matter. Section 3 contains some brief comments, in lieu of conclusions.\\


\section{Composition of entropies and induced generalized operations}

The $q-$ entropic functionals \cite{Ts1}  were defined/introduced in the Physics literature, in the case of discrete systems with probabilities \ $p_i$ \ 
 labelled by \ $i\in I$, \ where \ $I$ \  is an index set,  as
\begin{equation}
            \mathcal{S}_q [\{ p_i \}] \ = \  k_B \cdot  \frac{1}{q-1} \left(1 - \sum_{i\in I} p_i^q \right)
\end{equation}
where \ $k_B$ \ is the Boltzmann constant, occasionally set to one by an appropriate re-arrangement of units. In eq. (1) \ $q\in\mathbb{R}$ \  is a ``bias"/weight 
exponent called  ``entropic" or ``nonextensive" parameter. The generalization of eq. (1) to continuous systems, characterized by a probability distribution \ $\rho$ \ 
in their phase space \ $\Omega$ \  has always been assumed to be the straightforward expression 
\begin{equation}
         \mathcal{S}_q [\rho ] \ = \ k_B \cdot \frac{1}{q-1} \left(1 - \int_\Omega [\rho(x)]^q \ dvol_\Omega \right)
\end{equation}
where \ $dvol_\Omega$ \ indicates the infinitesimal volume element of \ $\Omega$. \ No serious justification was offered for eq. (2), so far as we know, until \cite{Abe1, LB}
raised some potentially serious shortcomings of this expression that would not only invalidate it, but would also question whether the $q$-entropies can be extended to 
continuous systems, or even be a valid entropic functionals, in general. This gave rise to the currently unsettled controversy, in our opinion, 
\cite{Abe1, Andresen, Abe2, BOT, LB, BL, PR1, OB1, OA, OB2} to which we would like to contribute.   \\ 

A question pertinent to eq. (2), is  what \ $\rho$ is, \  and how it can be determined. For ergodic systems and the BGS entropy, the answer is well-known: 
the infinitesimal measure of interest is the infinitesinal area of the constant energy hypersurface in phase space of an isolated system (the microcanonical distribution). 
 This area however is very rarely explicitly computable, hence the far more extensive use of the much more convenient canonical 
distribution, assumed to give equivalent resuts for the physically important quantities at least, to the microcanonical one, in the thermodynamic limit. 
But for non-ergodic actions, determining such a \ $\rho$ \ is not a priori so obvious, and it may be questionable whether it even exists, is well-defined, or whether it is unique.
 We pointed out in our previous work how such \ $\rho$ \  on \ $\Omega$ \ may arise from the dynamical evolution of the system \cite{NK1}. Moreover, we proposed to use a
local covariant quantity, the Bakry-\'{E}mery-Ricci ($N-$ Ricci) tensor, which determines how features of such systems \cite{NK2} may be  calculable in concrete cases, 
in principle at least. However, none of \cite{NK1,NK2} addressed the problem of the existence of such a \ $\rho$, \ but instead took the validity of eq. (2) for granted.\\ 

Following A. Einstein and L. Boltzmann, we believe that the existence of such a \ $\rho$ \ can only be determined  from the underlying dynamical system. 
Since this is not feasible, in the current state of  development of the subject or in the foreseeable future, we have to settle with less. 
 Part of a possible resolution of the existence, meaning, or validity, of eq. (2) may lie in the composition properties of \ $\mathcal{S}_q$. 
After all, this is the  most important difference between the BGS and the $q$-entropies, at least as detected from a comparison of 
their axiomatic formulations \cite{Santos, Abe3}.\\

One can readily see that the $q-$entropies, either in their discrete or in their continuous forms,  obey the composition property, with \ $k_B =1$,
\begin{equation}
   \mathcal{S}_q [\{ p_{A + B}\} ] \ =\ \mathcal{S}_q [\{ p_A \} ] + \mathcal{S}_q [ \{ p_B \} ]  + (1-q) \mathcal{S}_q [ \{ p_A \} ]  \mathcal{S}_q [ \{ p_B \} ] 
\end{equation}
for two independent systems \ $A, B$, \ which when interact they give rise to a system indicated by \ $A+B$, \ with 
probability distributions \ $p_A, \  p_B, \ p_{A+B}$ \ respectively. This motivated the introduction \cite{NMW, Borges} of the generalized addition   
\begin{equation}
            x\oplus_q y \ = \ x+y+ (1-q)xy, \ \ \  \ \ \ x,y \ \in \mathbb{R} 
\end{equation}
A generalized multiplication \ $\otimes_q$ \ which is distributive with respect to \ $\oplus_q$ \ was proposed in \cite{PCPB}, and in a different but conjecturally equivalent form in 
\cite{NK3}, to be 
\begin{equation}
   x\otimes_q y \ = \  \frac{(2-q)^{\frac{\log [1+(1-q)x] \cdot \log [1+(1-q)y]}{[\log (2-q)]^2}}  -1}{1-q}
\end{equation}
As a result,  a deformation of the reals \ $\mathbb{R}_q$ \  was proposed  in \cite{PCPB, NK3}, which is realized through the field isomorphism \
 $\tau_q: \mathbb{R}\rightarrow \mathbb{R}_q$ \  defined as
\begin{equation}  
      \tau_q (x) = \frac{(2-q)^x - 1}{1-q}
\end{equation}
This field isomorphism encodes the $q-$entropies' composition property eq. (3). We should point out  that eq. (4) is not the only generalized product that someone can come up with.
Indeed, the rather important \cite{STU} was derived based on the  generalized multiplication
\begin{equation}
   x\otimes_q y \ = \  \left[x^{1-q} + y^{1-q} - 1  \right]_+ ^\frac{1}{1-q}
\end{equation}
where \ $[x]_+ \ = \ \max \{x, 0 \}, \ x\in\mathbb{R}$.
The problem, as we see it, in adopting the generalized multiplication eq. (7)  is that it is not distributive with respect to the generalized addition eq.  (4) , 
something that impedes the construction of interesting generalized algebraic structures based on eq. (4), in our opinion. 
Hence, we will tacitly assume the use eq. (5) over eq. (7) in the sequel. \\

 As long as \ $0\leq q < 1$, \ with \ $q=1$ \ being the BGS entropy which is a special case of \ $\mathcal{S}_q$, \ 
there have been still some, but overall fewer, objections to the ``continuum limit" eq. (2) of eq. (1). Moreover, in this range, eq. (6) is well-defined and it is  indeed a field isomorphism. 
Outside this range of parameter values of $q$, things become less clear. One can refer to \cite{OB3, OB4} that address this issue. 
The conjectured dualities of \ $q$ \ may also provide extensions outside this range, but this is still
a subject of investigation.  The success of eq. (6) is corroborated by that it is possible, within its context, 
to define a generalized Fourier transform \cite{Scar, NK3} which is free of the problem of invertibility plaguing the other proposed  definitions  \cite{STU, PR2}.
The uniqueness of the generalized Fourier transform is guaranteed in the context of the generalized operations giving rise to eq. (6), by rigorous and well-established theorems. \\    

The definition of  the continuous version of,  even, the BGS entropy 
\begin{equation}
                         \mathcal{S}_{BGS} [\rho ] \ = \ - k_B \int_\Omega  \  \rho(x) \ \log \rho(x)  \ \ dvol_{\Omega}
\end{equation}
has some obvious drawbacks, which  although well-known,  may not as widely appreciated \cite{Balian} as they should be. One of them is that  it is not re-parametrizarion invariant. 
To restore such an invariance, one re-interprets the BGS functional as a relative entropy of the given,   with  respect to a reference distribution. In such a case 
$\rho$ in eq. (8) is not really a probability distribution, but rather the Radon-Nikod\'{y}m derivative of the probability measure  whose entropy we wish to calculate with respect to the 
reference measure. We tacitly assume during this process that these two measures are absolutely continuous with respect to each other, so  such a Radon-Nikod\'{y}m derivative exists
(almost) everywhere on \ $\Omega$.\ This does not create a conflict as the physically relevant quantity is the entropy change in classical Physics, rather than the entropy itself,
hence any contribution of the reference distribution in eq. (8) becomes irrelevant.  One can also mention at this point that  
 the Gibbsian definition eq. (8) gives the same result as Boltzmann's entropy only in the thermodynamic limit \cite{Jaynes1} even for 
simple systems. The above criticisms  in no way invalidate the use of the BGS entropy, or of the functional eq. (8),  for continuous systems. \\  
 
We believe that the $q$-entropies may deserve a similar treatment. Whether it is physically relevant, or not, and in what context, is a question that can only be decided experimentally. 
On theoretical grounds, we may have to 
rethink the validity of some even established concepts, if we want to have the freedom to accommodate the possbility of eq. (8) being actually valid for continuous systems.
We argued, for instance, in \cite{NK4} that if we accept the applicability of the $q$-entropies for a system of many degrees of freedom,  then the conventional, successful and 
time-honored definition of the Legendre-Fenchel transform has to be modified. This conclusion relied on model-indepedent statements, on the fundamental notion of convexity 
 and on the essence of what the Legendre-Fenchel transform really is.\\

Another issue may be how to properly define the thermodynamic limit for power-law entropies.  If the $q$-entropies indeed describe systems with long-range interactions, then how to 
consider the thermodynamic limit is even conceptually unclear. Such systems cannot effectively be put in a box whose  surface contributions can be ignored. On the contrary, 
such a boundary has a significant contribution to the entropy of the system. We cannot ignore such an artifical box by sending its lentgh to infinity in order to minimize its contribution
to the entropy. In  doing so, even for regularization purposes, we will get results that are highly suspect physically, and  may not even make sense mathematically.  \\  

What we propose is to use the generalized operations \ $\oplus_q$ \ and \ $\otimes_q$ \ stated in eq. (4) and eq. (5)  in such definitions. 
If we use these two operations to express the statistical  behavior of systems 
described by the $q$-entropies under composition, then these generalized operations should be used in the calculations of all thermodynamic quantities,
instead of the ordinary addition and multiplication. As a result, one would have to calculate for the entropy variations, either in the discrete or continuous case,  the quantity
\begin{equation}   
          \mathcal{S}_q [ \{ p_2 \} ] \ominus_q \mathcal{S}_q [ \{ p_1 \} ]
\end{equation}
where
\begin{equation}
          x\ominus_q y \ = \ \frac{x-y}{1+(1-q)y}
\end{equation}
is the generalized subtraction stemming from eq. (4), rather than the usual difference etc. \\

Following \cite{Jaynes2, Abe1},  consider \ $\Omega = [a,b] \subset \mathbb{R}$  \ sub-divided by \ $n-2$ \  interior points, \
\ $a=x_1 < x_2 < \ldots < x_{n-1} < x_n = b$, \ and  introduce the reference measure \ $m$ \ on \ $\Omega$ \ as
\begin{equation}
                 \frac{1}{m(x_i)} \ = \  \lim_{n\rightarrow\infty} n(x_{i+1} - x_i)
\end{equation}
Summation becomes integration in the transition from the discrete to the continuum case \cite{Abe1}
\begin{equation}
     p_i \ \rightarrow \ \frac{\rho(x_i)}{nm(x_i)}
\end{equation}
The argument of \cite{Abe1} is that in differences between \ $\mathcal{S}_{BGS}$ \ entropies
\begin{equation}   
    \lim_{n\rightarrow\infty}  \mathcal{S}_{BGS} [\{ p_n \} ] \ = \ - k_B \int_a ^b \ \rho(x) \ \log \frac{\rho(x)}{nm(x)} \  dx
\end{equation}
the regulating variable \ $n$ \ disappears due to the presence of the logarithm in eq.  (13). By contrast, this does not happen in differences between values of 
\  $\mathcal{S}_q$, \ which in this context is given by
\begin{equation}
     \lim_{n\rightarrow\infty}   \mathcal{S}_q [ \{ p_n \} ] \ = \ k_B \cdot \frac{1}{q-1}  \left\{1 - \int_a ^b \ \rho(x) \left[ \frac{\rho(x)}{nm_q(x)}\right]^{q-1}  - 1  \right\}
\end{equation}
So the limit  \ $n\rightarrow\infty$ \ does not really exist in differences of the $q-$entropies of the form of eq. (13). As a result 
eq. (13), and hence eq. (2),  does not have any physical  meaning and therefore it should be discarded.\\

Consider however the generalized subtraction eq. (10) for two values of eq. (14). Then the presence of \ $\mathcal{S}_q$ \  in the denominator of eq. (10) for either the discrete of the continuous forms of the $q$-entropies
\begin{equation}
     \mathcal{S}_q [\rho_2] \ominus_q \mathcal{S}_q [\rho_1] \ = \ \frac{\mathcal{S}_q [\rho_2] - \mathcal{S}_q [\rho_1]}{1+ (1-q) \mathcal{S}_q [\rho_1]}
\end{equation}
acts as a kind of regulator, so the result of such a difference is not necessarily infinite. \ Therefore an expression such as eq. (15) is not automatically
meaningless, cannot be discarded off-hand, but it has to be considered with greater care.   
The same idea should apply to the derivatives of \ $\mathcal{S}_q$, \ either in the discrete or in the continuous cases, which give rise to the different response functions such as capacities, 
susceptibilities etc: they should be determined with respect to \ $\oplus_q$ \ and \ $\otimes_q$ \ and the topologies, limits etc. induced by them, rather than through the ordinary 
operations of \ $\mathbb{R}$. \ What we essentialy state is that the $q$-entropies, and all non-BGS entropic functionals, in general, should be combined according to their generalized composition properties, which in the case of the $q$-entropies are \ $\oplus_q$ \ and \ $\otimes_q$, \ and not according to the usual binary operations of \ $\mathbb{R}$ \ which appear un-natural from the viewpoint of such entropies. Using such generalized operations may also help address the issue of how to consider the thermodynamic limit of systems described by
such  power-law entropies, as in the case of systems having long-range interactions, for instance.  The real difficulty, in our opinion, is to justify (or prove) the emergence of such generalized binary operations from the microscopic dynamics of such systems.\\

  In closing, we would like to stress that the above do not imply that the continuum limit of the $q$-entropies does exist, or that it does have some 
physical significance since the argument is completely formal. However,  it points out that a different viewpoint and a  
more careful treatment may be needed in  establishing the thermodynamic behavior of systems described by \ $\mathcal{S}_q$ \ compared to the ones described by 
\ $\mathcal{S}_{BGS}$.\\  


\section{Conclusions and discussion}

The authors believe that the issues raised in the standing controversy about the continuum limit and the meaning, if any, of the $q$-entropies are important but still not fully settled. 
We presented in this work an argument based on the generalized operations induced by the $q$-entropies' composition, which may be useful in addressing other issues pertinent to this debate
and to the use, if any, of the $q$-entropies in Physics. We strongly favor the extensive use of the generalized addition and multiplication in non-additive 
Thermodynamics and Statistical Mechanics, which phenomenological as they may be, reflect the macroscopic description of the combination of two subsystems. 
Ideally, we would like an actual derivation, even heuristic,  of such composition properties from the underlying dynamics of the system.\\

Our approach is similar in spirit to the requirement for a manifest way to express the invariances of physical theories,
such as the general covariance in General Relativity and Gauge theories, or  the covariance under supersymmetric transformations etc. 
The major difference of the above with our proposal, is that instead of changing the form of the equations by introducing appropriate covariant derivatives 
(connections) on vector buncles having the desired properties,  we propose to change the underlying binary operations of the fields defining such structures. 
In effect, our approach is closer to the spirit of modern Algebraic Geometry, mostly inspired by the influence of A. Grothendieck, where one considers algebraic 
equations over fields having common features \cite{Hart} rather than using the set of complex numbers \ $\mathbb{C}$ \ only.  This viewpoint has had limited 
effectiveness in Physics, so far as we know, but may turn out to have a  more promising future in the context of string/brane/M-theories \cite{BKOR}.\\  



\noindent{\bf Acknowledgement:}  \ \ We are grateful to the referees for their constructive criticism that helped improve the exposition, for suggesting several pertinent references  
and for their comments clarifying several points.  The second author is  grateful to Professor Anastasios Bountis  for his support, without which  this work
would have never been possible.  \\  






\begin{thebibliography}{99}
\bibitem{Ts1} C. Tsallis, \ \emph{Possible generalisation of boltzmann-Gibbs statistics}, \ J. Stat. Phys. {\bf 52}, \ 479-487 \ (1988).
\bibitem{Ts-book} C. Tsallis, \ \emph{Introduction to Nonextensive Statistical Mechanics: Approaching a Complex World}, \ Springer Science + Business Media, \ New York, NY, USA \  (2009). 
\bibitem{Kan}  G. Kaniadakis, \ \emph{Theoretical Foundations and Mathematical Formalism of the Power-Law Tailed Statistical Distributions}, \ Entropy {\bf 15}(10), \ 3983-4010 \ (2013).
\bibitem{CAT} L.J.L. Cirto, V.R.V. Assis, C. Tsallis, \ \emph{Influence of the interaction range on the thermostatistics of a classical many-body system}, \ Physica A {\bf 393}, 
                                           \ 286-296 \ (2014).
\bibitem{CRSA} G. Combe, V. Richefeu, M. Stasiak, A.P.F. Atman, \ \emph{Experimental validation of nonextensive scaling law in confined granular media}, \ Phys. Rev. Lett. {\bf 115}, \
                                                       238301 \ (2015).
\bibitem{AMPBIV} A. Argun, A.-R. Moradi, E. Pince, G.B. Bagci, A. Imparato, G. Volpe,   \ \emph{Non-Boltzmann stationary distributions and nonequilibrium relations in active baths}, \ Phys. Rev. E {\bf 94}, \ 062150 \ (2016).

\bibitem{BT} D. Bagchi, C. Tsallis, \ \emph{Long-ranged Fermi-Pasta-Ulam systems in thermal contact: Crossover from q-statistics for Boltzmann-Gibbs statistics}, \ Phys.  Lett. A {\bf 381}, \ 
                                                                         1123-1128 \ (2017).



\bibitem{Abe1} S. Abe, \ \emph{Essential discreteness in generalized thermostatistics with non-logarithmic entropy}, \ Europhys. Lett. {\bf 90}, \ 50004 \ (2010).
\bibitem{Andresen} B.B. Andresen, \ \emph{Comment on ``Essential discreteness in generalized thermostatistics with non-logarithmic entropy" by Abe Sumiyoshi}, 
                                                                 \ Europhys. Lett.  {\bf 92}, \ 40005 \ (2010).
\bibitem{Abe2} S. Abe, \ \emph{Reply to the comment by B. Andresen}, \ Europhys. Lett. {\bf 92}, \ 40006 \ (2010).
\bibitem{BOT} G.B. Bagci, T. Oikonomou, U. Tirnakli, \ \emph{Comment on ``Essential discreteness in generalized thermostatistics with non-logarithmic entropy" by S. Abe}, \
                                     {\sf arXiv:1006.1284 [cond-mat.stat-mech]} 
\bibitem{LB} J.P. Boon, J.F. Lutsko, \ \emph{Nonextensive formalism and continuous Hamiltonian systems}, \ Phys. Lett. A {\bf 375}, \ 329-334 \ (2011).
\bibitem{BL} J.F. Lutsko, J.P. Boon, \ \emph{Questioning the validity of non-extensive thermodynamics for classical Hamiltonian systems}, \ Europhys. Lett. {\bf 95}, \ 20006 \ (2011).
\bibitem{PR1} A. Plastino, M.C. Rocca, \ \emph{On the putative essential discreteness of q-generalized entropies}, \ Physica A {\bf 488}, \ 56-59 \ (2017).
\bibitem{OB1} T. Oikonomou, G.B. Bagci, \ \emph{Route from discreteness to the continuum for the non-logarithmic q-entropy}, \ Phys. Rev. E {\bf 97}, \ 012104 \ (2018).
\bibitem{OA} C. Ou, S. Abe, \ \emph{Comment on ``Route from discretness to the continuum for the Tsallis q-entropy"}, \ Phys. Rev. E {\bf 97}, \ 066101 \ (2018).
\bibitem{OB2} T. Oikonomou, G.B. Bagci, \ \emph{Reply to the comment on ``Route from discreteness to the continuum for the Tsallis q-entropy" by Conjie Ou and Sumiyoshi Abe}, \
                                 Phys. Rev. E {\bf 97}, \ 066102 \ (2018).
\bibitem{Gal} G. Gallavotti, \ \emph{Statistical Mechanics: A Short Treatise}, \ Texts Monog. Phys., \ Springer-Verlag, \ Berlin, Germany \ (1999).
\bibitem{Cohen} E.G.D. Cohen, \ \emph{Boltzmann and Einstein: Statistics and Dynamics - An unsolved problem}, \ Pramana {\bf 64}(5), \ 635-643 \ (2005).
\bibitem{NK1} N. Kalogeropoulos, \ \emph{Moduli of curve families and  (quasi-)conformality of power-law entropies}, \ Int. J. Geom. Methods Mod. Phys. 
                                                                           {\bf 13}, \ 1650063 \ (2016).
\bibitem{NK2} N. Kalogeropoulos, \ \emph{Ricci curvature, isoperimetry and a non-additive entropy}, \ Entropy {\bf 17}, \ 1278-1308 \ (2015).
\bibitem{Santos} R.J.V. Santos, \ \emph{Generalization of Shannon's theorem for Tsallis entropy}, \   J. Math. Phys. {\bf 38}, \ 4104-4107 \ (1997).  
\bibitem{Abe3} S. Abe, \ \emph{Axioms and uniqueness theorem for Tsallis entropy},    \  Phys. Lett. A {\bf 271}, \  74-79 \ (2000). 
\bibitem{NMW} L. Nivanen, A Le M\'{e}haut\'{e}, Q.A. Wang, \ \emph{Generalized algebra within nonextensive statistics}, \ Rep. Math. Phys. {\bf 52}, \ 437-444 \ (2003).
\bibitem{Borges} E.P. Borges, \ \emph{A possible deformed algebra and calculus inspired in nonextensive thermostatistics}, \  Physica A {\bf 340}, \ 95-101 \ (2004).
\bibitem{PCPB} T.C. Petit Lob\~{a}o, P.G.S. Cardoso, S.T.R. Pinho, E.P. Borges, \ \emph{Some properties of defomed q-numbers}, \ Braz. J. Phys. \ {\bf 39}, 402-407 \ (2009).
\bibitem{NK3} N. Kalogeropoulos, \ \emph{Distributivity and deformation of the reals from Tsallis entropy}, \ Physica A {\bf 391}, \ 1120-1127 \ (2012). 
\bibitem{STU}   S. Umarov, C. Tsallis, S. Steinberg, \ \emph{On a $q$-Central Limit Theorem Consistent with Nonextensive Statistical Mechanics}, \ 
                                                                    Milan J. Math. {\bf 76}(1), \  307-328 \ (2008). 
\bibitem{OB3} T. Oikonomou, G. B. Bagci, \ \emph{The maximization of Tsallis entropy with complete deformed  functions and the problem of constraints}, \  
                                                                    Phys. Lett. A {\bf 374}, \  2225-2229 \  (2010). 
\bibitem{OB4} G. B. Bagci, T. Oikonomou, \ \emph{Validity of the third law of thermodynamics for the Tsallis entropy}, \ Phys. Rev. E {\bf 93}, \ 022112 \ (2016). 

\bibitem{Scar} A.M. Scarfone, \ \emph{$\kappa$-deformed Fourier transform}, \ Physica A {\bf 480}, \ 63-78 \ (2017).
\bibitem{NK3} N. Kalogeropoulos, \ \emph{The $\mathrm{\tau}_q$ Fourier transform: covariance and uniqueness}, \ Mod. Phys. Lett. B  {\bf 32}(14), \ 1850149 \ (2018).
\bibitem{PR2} A. Plastino, M.C. Rocca, \ \emph{Inversion of Tsallis' $q$-Fourier transform and the complex plane generalization},  \ Physica A {\bf 391}, \ 4740-4747 \ (2012).
\bibitem{Balian}  R. Balian,  \ \emph{Entropy: a Protean Concept}, \ S\'{e}m. Poincar\'{e} {\bf 2}, \ 13-27 \ (2003).
\bibitem{Jaynes1} E.T. Jaynes, \ \emph{Gibbs vs Boltzmann Entropies}, \ Amer. J. Phys. {\bf 33}(5), \ 391-398   \ (1965). 
\bibitem{NK4} N. Kalogeropoulos, \ \emph{The Legendre Transform in Non-Additive Thermodynamics and Complexity}, \ Entropy {\bf 19}(7), \ 298 \ (2017). 
\bibitem{Jaynes2} E.T. Jaynes, \ \emph{Probability Theory: The Logic of Science}, \ Cambridge Univ. Press, \ Cambridge, UK (2003). 
\bibitem{Hart} R. Hartshorne, \ \emph{Algebraic Geometry}, \  Grad. Texts Math., \ Springer Science + Business, \ New York, NY, USA \ (1977).
\bibitem{BKOR} N. Benjamin, S. Kachru, K. Ono, L. Rolen, \ \emph{Black holes and class groups}, \ {\sf arXiv:1807.00797 [math.NT]}  
\end{thebibliography}
\end{document}